\documentstyle[12pt]{article}

\begin{document}
{\sf \begin{center} \noindent
{\Large \bf Plasma metric singularities in helical devices and tearing instabilities in tokamaks}\\[3mm]

by \\[0.3cm]

{\sl L.C. Garcia de Andrade}\\

\vspace{0.5cm} Departamento de F\'{\i}sica
Te\'orica -- IF -- Universidade do Estado do Rio de Janeiro-UERJ\\[-3mm]
Rua S\~ao Francisco Xavier, 524\\[-3mm]
Cep 20550-003, Maracan\~a, Rio de Janeiro, RJ, Brasil\\[-3mm]
Electronic mail address: garcia@dft.if.uerj.br\\[-3mm]
\vspace{2cm} {\bf Abstract}
\end{center}
\paragraph*{}
Plasma toroidal metric singularities in helical devices and
tokamaks, giving rise to magnetic surfaces inside the plasma devices
are investigated in two cases. In the first we consider the case of
a rotational plasma on an helical device with circular cross-section
and dissipation. In this case singularities are shown to place a
Ricci scalar curvature bound on the radius of the surface where the
Ricci scalar is the contraction of the constant Riemannian curvature
tensor of magnetic surfaces. An upper bound on the initial magnetic
field in terms of the Ricci scalar is obtained. This last bound may
be useful in the engineering construction of plasma devices in
laboratories. The normal poloidal drift velocity is also computed.
In the second case a toroidal metric is used to show that there is a
relation between singularities and the type of tearing instabilities
considered in the tokamak. Besides, in this case Ricci collineations
and Killing symmetries are computed.The pressure is computed by
applying these constraints to the pressure equations in tokamaks.
\vspace{0.5cm} \noindent {\bf PACS numbers:}
\hfill\parbox[t]{13.5cm}{02.40.Hw-Riemannian geometries}

\newpage
\section{Introduction}
 Geometrical techniques have been used with great success \cite{1} in Einstein general relavity have been also used
 in other important areas of physics, such as plasma structures in tokamaks as been clear in the seminal book by Mikhailovskii \cite{2}
 to investigate the tearing and other sort of instabilities in confined plasmas \cite{3}, where the Riemann metric
 tensor plays a dynamical role interacting with the magnetic field through the magnetohydrodynamical equations (MHD). Recentely
 Garcia de Andrade \cite{4} has applied Riemann metric to investigate magnetic flux tubes in superconducting plasmas.
 Thiffault and Boozer \cite{5} have also applied the methods of Riemann geometry in the context of chaotic flows and fast
 dynamos. In this paper we use the tools of Riemannian geometry, also used in other branches of physics as general relativity
 \cite{6}, such as Killing symmetries , Riemann and Ricci \cite{7} collineations , shall be applied here to generate
 magnetic nested surfaces in helical devices and tokamaks through the built of Einstein spaces obtained from Tokamak plasma metric. This work is motivated
 by the fact that the magnetic surfaces are severely constrained in tokamaks \cite{3} and Killing symmetries are well applied every time we have symmetries in the problem as in
 solutions of Einstein equations of general relativity. Equilibrium of these surfaces or their instabilities are fundamental
 in the constructio of tokamaks and other plasma devices such as stellarators where torsion is also present. The magnetic surfaces
 are more easily obtained when symmetries are present. This is our main motivation to apply the special Riemann geometrical techniques
 of Killing symmetries and Ricci collineation to obtain magnetic surfaces formed by Einstein spaces. To simplify matters we shall consider
 two usual approximations from plasma physics \cite{3} which are the small toroidality or inverse aspect ratio ${\epsilon}=\frac{a}{R}<<1$ where here R represents the external
 radius of the torus and a is its internal radius, and the Shafranov displacement ${\Delta}'<<1$ as well. We consider examples
 of two plasma metrics: The first is the metrical of plasma rotation in tokamaks where plasma dissipation is taken into account.
 This metric was used for the first time by Tsypin et al \cite{8} to describe dissipative plasmas in the circular cross-section
 helical devices such as HELIACS or Drakon (stellarator) \cite{9}. The pressure of the tokamak is obtained from the tokamak Shafranov shift equation.
 Constant pressure closed ergodic nested surfaces in magnetohydrostatics  have also been shown
 by Schief \cite{10} to be generated by solitons. This is another mathematical technique , distinct from ours, is another use ofmathematical theory to generate of nested magnetic surfaces in
 plasmas. The paper is organised as follows:
 In section 2 we review the Ricci tensor techniques and Ricci colineations which are not usually familiar
 to the plasma physicists. In section 3 we solve the Ricci tensor components from the plasma metric by considering
 that nested surfaces are formed by Einstein spaces, where the Ricci tensor is proportional to the metric. In section 4
 stablishing a geometrical method for the classification of tearing instabilities and solve the
 Ricci collineation equations to find out the Killing vectors for the dissipative rotational plasma metric..
 Conclusions are presented in section 5.
 \section{Ricci collineations from plasma metrics}
 Ricci tensor $R_{ik}$ is contructed from the contraction of Riemann
 tensor $R_{ijkl}$ in terms of the contravariant components of the metric by the expresion
\begin{equation}
 R_{ik}=g^{jl}R_{ijkl}\label{1}
\end{equation}
and the Ricci scalar R by an extra contraction in terms of the
metric tensor as $R=g^{ij}R_{ij}$. Let us now compute the Riemann
space of constant curvature represented by the Riemann tensor
components
\begin{equation} R_{ijkl}={\Lambda}(g_{ik}g_{jl}-g_{il}g_{jk}) \label{2}
\end{equation}
where ${\Lambda}$ is a constant which is called de Sitter
cosmological constant. Contraction of expression (\ref{2}) in two
non-consecutive indices,otherwise the symmetry of the Riemann
curvature tensor $R_{ijkl}= -R_{jikl}=R_{jilk}$ would make them
vanish, yields the Einstein space Ricci relation
\begin{equation}
R_{ik}=2{\Lambda}g_{ik}\label{3}
\end{equation}
 The Ricci collineations equations are given by
\begin{equation}
[{\partial}_{l}R_{ik}]{\eta}^{l}+R_{il}{\partial}_{k}{\eta}^{l}+R_{kl}{\partial}_{i}{\eta}^{l}=0
\label{4}
\end{equation}
where ${\eta}^{l}$ are the components of the Killing vector
$\vec{\eta}$ which defines the symmetries of the associated space,
and ${\partial}_{l}:=\frac{{\partial}}{{\partial}x^{l}}$ are the
components of the partial derivative operator. This equation is
obtained from the more elegant definition in terms of the Lie
derivative ${\cal L}_{\eta}$ as
\begin{equation}
{\cal L}_{\eta}R_{ik}=0 \label{5}
\end{equation}
In the next section we shall construct the magnetic surface as
plasma metric singularities and in section IV we solve the Ricci
collineation equations in terms of the plasma metric above.
\section{Rotational Plasmas Metric Singularities in Helical Devices}
 Let us now consider the application of the Tsypin et al metric of a
 rotational toroidal dissipative plasma tokamak, given by the
 nonvanishing components
 \begin{equation}
g_{11}=\frac{1}{4{\pi}B_{0}{\phi}} \label{6}
\end{equation}
\begin{equation}
g_{22}= \frac{\phi}{{\pi}B_{0}}\label{7}
\end{equation}
\begin{equation}
g_{23}=R {\tau}_{0}\frac{\phi}{{\pi}B_{0}}\label{8}
\end{equation}
\begin{equation}
g_{13}=\frac{1}{2{\pi}{B_{0}}^{\frac{1}{2}}}
{\partial}_{\eta}(\frac{1}{{B_{0}}^{\frac{1}{2}}}) \label{9}
\end{equation}
\begin{equation}
g_{33}={R^{2}}[1-(\frac{\phi}{{\pi}B_{0}})^{\frac{1}{2}}[Kcos{\theta}]]
\label{10}
\end{equation}
where the coordinates $(a,{\theta},{\eta})$ are respectively the
internal radius of the torus, which is constant on the magnetic
surface, and the remaining coordinates are the poloidal and toroidal
angles. The curvature ${\kappa}$ of the magnetic axis depends in
general of toroidal coordinate, and the torsion of the magnetic axis
is given by ${\tau}$. In this paper a great simplification would be
possible by considering that the torsion and curvature would be
constant. These metric allowed Tsypin et al \cite{8} to consider ion
viscosity in the plasma. Further ahead we shall also consider the
computation of the velocity in terms on the singular metric magnetic
surface. The magnetic field which is chosen in the form
$B=B_{0}[1-{\eta}_{t}cos{\theta}-{\eta}_{h}cos(m{\theta}-n{\eta})]$,
where ${\eta}_{t}$ is the toroidal amplitude of the magnetic field
spectrum and ${\eta}_{h}$ is the helical amplitude, while $(m,n)$
are the poloidal and toroidal modes of the helical magnetic field.
Here we shall consider that the $B_{0}$ is constant and that the
metric component $g_{23}\cong{0}$ due to to torsion weakness
assumption. Constancy of $B_{0}$ also yields $g_{13}$ also vanishes
, which turns the plasma helical metric diagonal. Let us now
computing the Riemann tensor component
\begin{equation}
R_{1212}=+\frac{{\partial}^{2}g_{12}}{{\partial}x^{1}{\partial}x^{2}}-\frac{{\partial}^{2}g_{11}}{{\partial}x^{2}{\partial}x^{2}}
-\frac{{\partial}^{2}g_{22}}{{\partial}x^{1}{\partial}x^{1}}
\label{11}
\end{equation}
which yields for the above metric
\begin{equation}
R_{1212}=\frac{{\phi}"}{{\pi}B_{0}} \label{12}
\end{equation}
Since we are looking for nested magnetic surfaces which possess
constant curvature the helical metric above must satisfy the
relation
\begin{equation}
R_{ijkl}={\Lambda}(g_{ik}g_{jl}-g_{ij}g_{kl}) \label{13}
\end{equation}
which in our case yields
\begin{equation}
R_{1212}={\Lambda}(g_{11}g_{22}-g_{12}g_{12}) \label{14}
\end{equation}
or
\begin{equation}
R_{1212}=\frac{{\Lambda}}{4{\pi}^{2}{B^{2}}_{0}} \label{15}
\end{equation}
Since the curvature component $R_{1212}$ is the same in both
expressions (\ref{12}) and (\ref{15}), equating them results in
\begin{equation}
{{\phi}"}=\frac{{\Lambda}}{4{\pi}B_{0}} \label{15}
\end{equation}
By integration one obtains
\begin{equation}
{{\phi}}=\frac{{\Lambda}a^{2}}{16{\pi}B_{0}}+c_{1}a+c_{2} \label{16}
\end{equation}
where $c_{1}$ and $c_{2}$ are integration constants. Now as usual in
general relativity we assume that a singularity could be obtained
either by the equations $g_{11}=\infty$ and $g_{22}=0$, as happens
as in Schwarzschild static black hole metric solution of Einstein
vacuum field equation. This set of equations can be obtained from
the helical metric above by setting ${\phi}(a)=0$, which from
(\ref{16}) yields the following second order algebraic equation in
the radius a
\begin{equation}
\frac{{\Lambda}a^{2}}{16{\pi}B_{0}}+c_{1}a+c_{2}=0 \label{17}
\end{equation}
which solution is
\begin{equation}
{a_{0}}^{\pm}=
\frac{{8{\pi}B_{0}}}{\Lambda}[c_{1}(-1\pm[1-\frac{\Lambda}{8{\pi}B_{0}}])]
\label{18}
\end{equation}
Thus we are left with two solutions , the simpler of which is
${a_{0}}^{+}=-c_{1}$ and
\begin{equation}
{a_{0}}^{-}=[1-\frac{{16{\pi}B_{0}}}{\Lambda}]c_{1}\label{19}
\end{equation}
Note that if $c_{1}>0$ solution ${a_{0}}$ is unphysical because the
radius of the helical plasma device cannot be negative, however the
physical solution ${a_{0}}^{-}$ yields a magnetic field bound from
the Ricci scalar
\begin{equation}
B_{0}\le\frac{{\Lambda}c_{1}}{16{\pi}{c_{2}}^{2}}\label{20}
\end{equation}
To obtain the Ricci collineations we compute first the Ricci scalar
components, since the metric is diagonal the only surviving
components are
\begin{equation}
R_{11}=g^{22}R_{l212}=\frac{\Lambda}{4{\phi}{\pi}B_{0}}\label{21}
\end{equation}
\begin{equation}
R_{22}=g^{11}R_{l212}= {\Lambda}{\phi}\label{23}
\end{equation}
These equations yields the Ricci scalar as
\begin{equation}
R=g^{11}R_{11}+g^{22}R_{22}={\Lambda}(1+{\pi}B_{0}){\phi}\label{24}
\end{equation}
since in general in plasma devices the field $B_{0}$ is very strong
we can consider that ${{\pi}{B^{2}}_{0}}>>B_{0}$ which from
expressions (\ref{20}) and (\ref{24}) yields an expression which
bounds $B_{0}$ in terms of the Ricci curvature scalar as
\begin{equation}
B_{0}\le\frac{\sqrt{c_{1}R}}{4{\pi}{c_{2}}} \label{25}
\end{equation}
This formula can certainly help in the building of new stellarators
and toroidal plasma devices in general.
\section{Magnetic surface singularities in tokamaks and Ricci collineations}
 Let us now start by considering the plasma metric given by Zakharov and Shafranov \cite{11} to investigate the evolution of
 equilibrium of toroidal plasmas. The components $g_{ik}$
 (i,k=1,2,3) and $(a,{\theta},z)$ as coordinates, of their plasma metric are
\begin{equation}
g_{11}=1-2{\Delta}'cos{\theta}+{{\Delta}'}^{2} \label{26}
\end{equation}
\begin{equation}
g_{22}=a^{2} \label{27}
\end{equation}
\begin{equation}
g_{33}=(R-{\Delta}+acos{\theta})^{2} \label{28}
\end{equation}
\begin{equation}
g_{12}=a{\Delta}'sin{\theta} \label{29}
\end{equation}
where $z=asin{\theta}$ and the dash represents derivation with
respect to a. In our approximation the last term in the expression
(\ref{1}) may be dropped. The magnetic surface equations are tori
with circular cross-section and equations
\begin{equation}r= R-{\Delta}(a)+acos{\theta} \label{30}
\end{equation}
\begin{equation}
z=asin{\theta} \label{31}
\end{equation}
 We shall consider now just two independent coordinates
$(x^{1}=a,x^{2}={\theta})$ since nested surfaces are bidimensional
in the case of plasmas, Let us now compute the Riemann tensor
components in the linear approximation
\begin{equation}
R_{1212}=+\frac{{\partial}^{2}g_{12}}{{\partial}x^{1}{\partial}x^{2}}-\frac{{\partial}^{2}g_{11}}{{\partial}x^{2}{\partial}x^{2}}
-\frac{{\partial}^{2}g_{22}}{{\partial}x^{1}{\partial}x^{1}}
\label{32}
\end{equation}
Substitution of the plasma metric above into expression (\ref{32})
yields the expression
\begin{equation}
R_{1212}=[3{\Delta}'cos{\theta}-2] \label{33}
\end{equation}
It is easy to show that the components $R_{1313}$ and $R_{2323}$
both vanishes within our approximations.  At this point we consider
that ${\theta}$ is so small that $sin{\theta}$ vanishes and
$cos{\theta}=1$ this simplifies extremely our metric and turns it
into a diagonal metric where $g_{12}=0$ and $g^{bb}=(g_{bb})^{-1}$
$(b=1,2)$ and this allows us to compute the components of the Ricci
tensor from the Riemann component. But before that let us compute
the use the condition that the nested surface is an Einstein space
to compute the Riemann component again
\begin{equation}
R_{1212}={\Lambda}a^{2}[1-2{\Delta}'cos{\theta}] \label{34}
\end{equation}
Since both expressions for the Riemann component $R_{1212}$ must
coincide, equating expressions (\ref{33}) and (\ref{34}) yields an
expression for the derivative of the Shafranov shift ${\Delta}$ as
\begin{equation}
{\Delta}'=-6[1-\frac{{\Lambda}a^{2}}{6}] \label{35}
\end{equation}
Integration of this expression yields the value of the shift in
terms of the radius a as
\begin{equation}
{\Delta}=-6a(1+\frac{{\Lambda}a^{2}}{12}] \label{36}
\end{equation}
which satisfies the well-known boundary condition ${\Delta}(0)=0$.
From these expressions one may also compute ${\Delta}"=2{\Lambda}a$.
An important result in plasma physics is that tearing instabilities
coming from ion or electron currents possess the shift condition
${\Delta}'<0$. This condition would be clearly fulfilled from
expression unless the ${\Lambda}$ curvature constant would be
negative and in modulus $\frac{{\Lambda}a^{2}}{2}< -1$. Note that
this situation is very similar to the condition of favorable or
unfavorable curvature for the instabilities in plasmas \cite{3}. The
main difference is that here we are refereeing to Riemann curvature
and not to Frenet curvature of the magnetic lines in plasmas. This
suggests another method to classify geometrically tearing
instabilities. Actually, since has been shown \cite{12} recently
that the Riemann tensor in plasmas can be expressed in terms of the
Frenet curvature both methods seems to be equivalent. Now let us
compute the Ricci components $R_{11}$ and $R_{22}$ from the
component $R_{1212}$ by tensor contraction with metric components
$g^{11}$ and $g^{22}$, which results in the expressions
\begin{equation}
R_{11}={\Lambda}[1-2{\Delta}'] \label{37}
\end{equation}
and
\begin{equation}
R_{22}=-[2+{\Delta}']\label{38}
\end{equation}
which in turn yields the expressions
\begin{equation}
{\partial}_{1}R_{11}=-2{\Lambda}{\Delta}" \label{39}
\end{equation}
and
\begin{equation}
{\partial}_{1}R_{22}=-{\Delta}"\label{40}
\end{equation}
From equations for $i=1,k=2$ one obtains
\begin{equation}
{\partial}_{1}{\eta}_{2}=0\label{41}
\end{equation}
\begin{equation}
{\partial}_{2}{\eta}_{1}=0\label{42}
\end{equation}
Substitution of these derivatives of the Ricci tensor components
into the Ricci collineations equations one obtains the following set
of PDE equations
\begin{equation}
2[2+{\Delta}']{\partial}_{1}{\eta}^{1}+{\eta}^{1}{\Delta}"=0
\label{43}
\end{equation}
which yields
\begin{equation}
{\eta}^{a}=[1+{\Delta}'] \label{44}
\end{equation}
Due to constraint $(\ref{24})$ the only solution for the equation
\begin{equation}
-[2+{\Delta}']{\partial}_{2}{\eta}^{2}-(1+{\Delta}'){\Delta}"=0
\label{45}
\end{equation}
is ${\eta}^{2}=0$.This allows us finally to write down the Killing
vector as
\begin{equation}
\vec{\eta}=[(1+{\Delta}'),0]\label{46}
\end{equation}
the last relation which this vector will have to satisfy
$g_{ij}({\eta}^{j})^{2}=1$ for the modulus of the Killing vector
will allow us to determine ${\Delta}$ and in turn from expression
(\ref{16}) will allow us to determine the nested surface radius $a$
in terms of the curvature constant ${\Lambda}$. This  implies that
\begin{equation}
{\Delta}'=\frac{3}{10}\label{47}
\end{equation}
or ${\Delta}=\frac{3}{10}a$, which satisfies the well-known boundary
condition ${\Delta}(0)=0$. Expression (\ref{29}) yields that
${\Delta}'>0$ which shows physically that the tearing instability
cannot come from ion or electron currents \cite{2}. Substitution of
this result into (\ref{16}) yields
\begin{equation}
a_{0}=\sqrt{\frac{19}{15}{\Lambda}} \label{48}
\end{equation}
The pressure now is easily computed from the expression given in
reference $3$
\begin{equation}
-2{\pi}c^{2}p\frac{a^{4}}{R}=-6aJ^{2}{\Delta}-(1-\frac{a}{R})J^{2}\label{49}
\end{equation}
where by using the value of  ${\Delta}$ yields
\begin{equation}
p=\frac{{R}J^{2}}{3{\pi}c^{2}{\Lambda}^{2}}\label{50}
\end{equation}
one notes that a singularity in the pressure decreases as the
curvature constant increases which agrees with the reasoning that
\cite{11} curvature tends to stabilize the plasma.
\section{Conclusions}
 In conclusion, we have investigated a method of classification and identification of tearing instability,
 allowing for example to distinguish between tearing instabilities that comes from ions and electron currents or
 not, based on the Riemann curvature constant submanifolds as nested
 surfaces in Einstein spaces. The Killing symmetries are shown also
 to be very useful in the classification of plasma metrics in the
 same way they were useful in classifying general relativistic
 solutions of Einstein's gravitational equations in four-dimensional
 spacetime \cite{6}. Since as it is well-known \cite{3} the ${\Delta}'$ behaves as $\frac{1}{{\delta}W}$ near marginal stability, we must
 conclude that there is a relation between the stability ${\delta}W>0$ or instability ${\delta}W<0$ and the positive
 or negative Riemann curvature of the nested surfaces discussed here. Other interesting examples of the utility of
 the is method is the Ricci collineations investigations of the
 twisted magnetic flux tubes and the Arnolds metric for the fast
 dynamo \cite{13,14,15}. Though the examples worked here keep some
 ressemblance to analog gravity models \cite{16} since our black
 hole analogy does not carry much to the metric since the plasma
 metric we use is a real plasma metric and not a pseudo-Riemannian
 plasma metric built from the scalar wave equation.
 \section*{Acknowledgements}
 Thanks are due to CNPq and UERJ for financial supports.

\end{document}